\documentstyle[psfig]{aa}

\begin{document}
\thesaurus{1()}
\title{Fundamental properties of the new dwarf galaxy And~VI - alias \lq Pegasus Dwarf\rq\ - another companion of M31}
\titlerunning{The And~VI dwarf galaxy}
\author{Ulrich Hopp\inst{1}, Regina E. Schulte-Ladbeck\inst{2}, Laura Greggio\inst{1}\inst{3}, D\"orte Mehlert\inst{1}
\thanks{Visiting Astronomer, German-Spanish Astronomical Center, Calar
Alto, operated by the Max-Planck-Institut f\"ur Astronomie jointly
with the Spanish National Commission for Astronomy}
}
\offprints{U. Hopp}
\institute{\inst{1} Universit\"atssternwarte M\"unchen, 
Scheiner Str. 1, D 81679 M\"unchen, FRG, email: hopp@usm.uni-muenchen.de\\
\inst{2} University of Pittsburgh, Pittsburgh, PA 15260, USA, 
email: rsl@phyast.pitt.edu\\
\inst{3} University of Bologna, I 40100 Bologna, Italy
}
\authorrunning{Hopp et al.}
\maketitle
\begin{abstract}

We present medium deep CCD imaging in B, V, and I of the Pegasus Dwarf
galaxy (And~VI) which was recently found by Karachentsev \&
Karachentseva (1998), and independently also by Armandroff et
al. (1999). The Calar Alto 2.2m images show a low surface brightness
galaxy. Its structure resembles that of the other known dSph
companions of M31 And~I,~II,~III, and V. The brightest stars are
resolved in all three colors. Color-magnitude diagrams in either B$-$V
or V$-$I show the tip of the red giant branch which allows us to estimate
a true distance modulus of $24.5 \pm 0.2$. The color-magnitude
diagrams and the structure show no evidence for recent star formation,
thus, a classification as spheroidal dwarf galaxy with a rather old
population seems appropriate. The total absolute magnitude of this
dwarf is M$_{V,0} = -10.4\pm0.2$.

\end{abstract}

\thesaurus{1(fill in)}

\section{Introduction}

Until recently, only three dwarf spheroidal (dSph) companions of M31
were known, namely And I, II, and III while no less than 9 such
systems are associated with the Milky Way Galaxy (MWG). The M31
companions have been known since 1972 and were found by a visual
inspection of the (old) Palomar Sky Survey by van den Bergh
(1972). Recently, new searches were initiated based on the new, deeper
plates of the POSS II. Armandroff et al. (1998) found and confirmed a
new M31 companion which they named And~V. Karachentsev \&
Karachentseva (1998) announced the finding of two low surface
brightness objects, namely the Cassiopeia Dwarf of B$\sim$16 and the
Pegasus Dwarf of B$\sim$14.5. The Pegasus dwarf was meanwhile found
independently by Armandroff et al. (1999), they called it And
VI. According to the location in the sky and the morphology, both
newly identified dwarf candidates might be additional M31
companions. As argued by Armandroff et al., the new survey plates are
not only more sensitive than those available to van den Bergh, but
were searched also to larger M31-centric distances, since the MWG
companions implied larger dimensions of the companion systems.  We
here present ground-based CCD observations done with the Calar Alto
2.2m telescope of the later candidate in constellation Pegasus and will
show that it is indeed a dSph companion to M31. Therefore, one might
prefer Armandroff's et al. designation as And~VI. In the following, we
will use this later name which refers (as the name Pegasus dwarf) to a
faint, low surface brightness object at RA 23$^h$ 49.2$^m$ DEC +24$^o$
19' (1950). The name And~VI avoids confusion with the Pegasus irregular dwarf
galaxy and outlines that the new dwarf galaxy belongs to the M31 system.

The observations are briefly presented in section 2 of this letter and
discussed in section 3. We conclude on the distance and thus on the
satellite nature of the And~VI galaxy in section 4, while all other
aspects which can be derived from our data are postponed to a later
paper.

\section{Observations and Reductions}

The Calar Alto 2.2m telescope and its focal reducer Cafos was used in
August 1998 to obtain images in B, V, and I. We obtained 3 frames in
every band, slightly shifted from one exposure to the next. Total
exposure times were 1800, 1500, and 1500 sec, respectively. The
observations were accompanied by the usual CCD calibration measurements
for flat-fielding and de-biasing. The flux calibration was established
by observing two Landolt (1992) faint standard fields several times at
various airmass. The seeing was 1.5 (I), 1.5 (V), and 1.7 (B) arc sec
(FWHM).  An I fringe pattern frame was derived from all 24 science
frames of this run with various locations on the sky. This fringe
pattern frame was substracted after debiasing and flatfielding, while
the B and V frames were only debiased and flatfielded. The field of
view of the frames is 12 by 12 arcmin, far larger than the extent of
the And~VI dwarf galaxy, estimated as 2 by 4 arc min by
Karachentsev \& Karachentseva.

The co-added And~VI frames were searched for stellar
objects in each band independently by the MIDAS 97NOV version of
Daophot (Stetson, 1987). The PSF was constructed from some 30 to 40
brighter, non-saturated, and well isolated stars. The PSF-photometry
was done with the ALLSTAR routine. Several thousand objects were
detected in each of the bands.  Aperture photometry of the standard
stars allowed us to calibrate the PSF magnitudes.  We are using here
only those stars which are detected in at least two colors, either V
and I, or B and V. This largely reduces false detections. 794 stars
were simultaneously identified in B and V while 979 were found in V and
I. The PSF photometric errors become larger than 0.15 at 23.75
(B), 23.25 (V), and 21.75 (I). The faintest objects detected are
typically a magnitude fainter than this, but we will use in the
following only those stars with errors less than 0.15.

As we here intend only to describe the dwarf galaxy and to establish
its distance from the location of the tip of the red giant branch
(TRGB), we postpone artificial star tests to evaluate the
incompleteness to a later paper.

To get a quantitative description of the structure of 
And~VI, we cleaned the V frame from the brightest stars (about
18, they are probably MWG foreground) and smoothed the images
with a large spatial filter (11 by 11 arc sec). On this smoothed
frame, we derived the structural parameters by applying the ellipticity
fit of Bender \& M\"ollenhoff (1987).

\section{Results and Discussion}

\begin{figure} 
\centerline{\psfig{figure=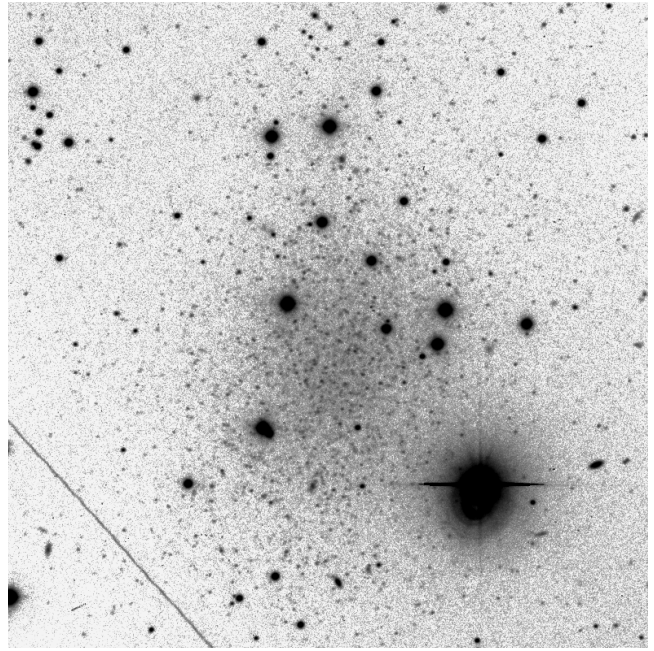,width=8.5cm,clip=t}}
\caption[]{Calar Alto 2.2m telescope CCD image in V of the Pegasus
dwarf galaxy = And~VI. North top, East left, field of view shown is
399.5 arc sec each side.}
\end{figure}

Figure 1 shows the V frame of the And~VI dwarf galaxy. It is
clearly resolved into individual, faint stars. Similarly, the B and I
band frames show faint, resolved stars. The images resemble those of
the other M31 dSph companions (see e.g. And~V in Armandroff et al.,
1998). The stars are very regularly distributed and show no structure
or knots which might indicate recent star formation as in dwarf
irregular galaxies. From the smoothed frame, we measured a central
surface brightness $\mu(0)$ in V of $24.24\pm0.1$ mag/$\Box"$ (24.13
when corrected for galactic reddening following Burstein \& Heiles,
1978). The light distribution follows nicely an exponential law of
scale length $79\pm3$ arc sec and has an overall ellipticity of
$0.20\pm0.04$ (Fig. 2). On the same images, we measured the total
magnitude inside the 25.5 mag/$\Box"$ V isophote (which essentially
corresponds to the Holmberg radius) of V = $14.17\pm0.2$ (already
corrected for a galactic reddening).  These data indicate a
classification as dSph.

\begin{figure} 
\centerline{\psfig{figure=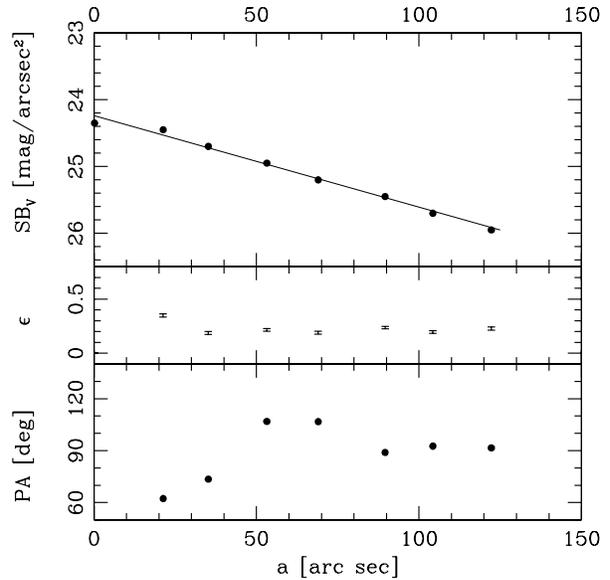,width=8.5cm,clip=t}}
\caption[]{Results of the surface photometry on the V frame. Top:
Surface brightness versus major axis $a$, the line shows the
fit. Middle: Ellipticity of the isophotes versus $a$. Bottom: Position
angle versus $a$.}
\end{figure}

\begin{figure} 
\centerline{\psfig{figure=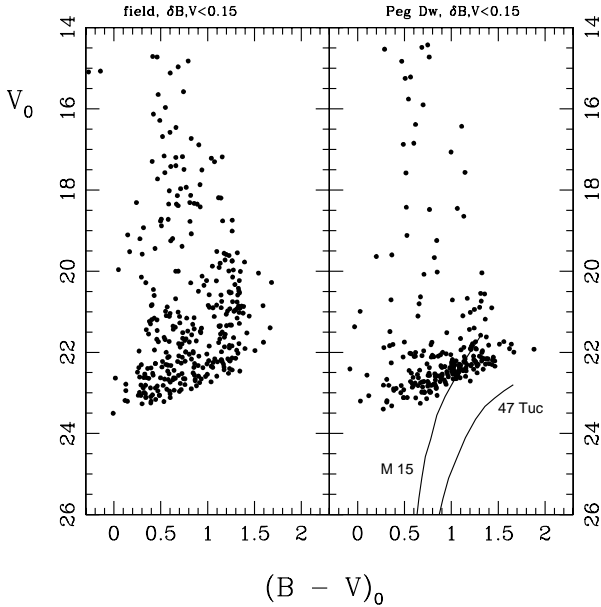,width=8.5cm,clip=t}}
\caption[]{V, B$-$V color magnitude diagram. Left: Stars in the field
surrounding And~VI, illustrating the MWG contamination. Right: Stars
either belonging to the And~VI dwarf galaxy or MWG stars being
projected on it. The left panel contains stars from the rest of the
CCD frame, an area 3.5 times larger than the area of the dwarf
galaxy. Both CMDs are corrected for galactic foreground reddening of
0.15 mag. in B. The right panel also shows as solid lines the red
giant branches for M15 from Durrel \& Harris (1993), a very low metal
abundance globular cluster, and for 47 Tuc from Hesser et al. (1987),
a high metal abundance globular. The RGB's of the two MWG globulars
have been shifted to a distance modulus of 24.5.}
\end{figure}

\begin{figure} 
\centerline{\psfig{figure=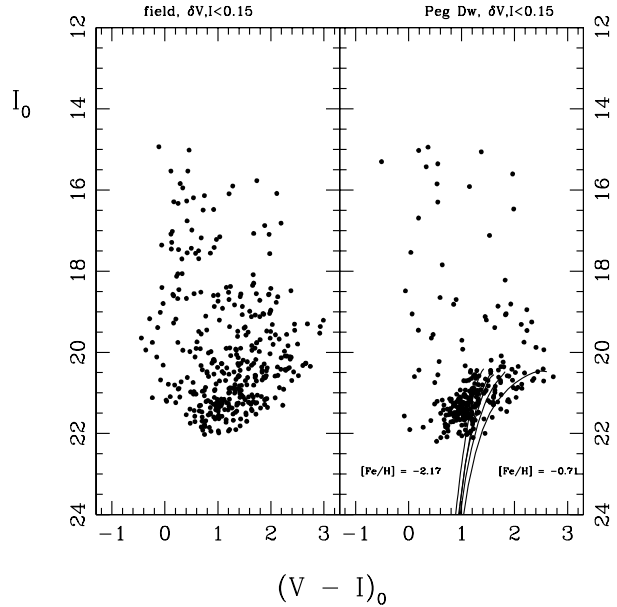,width=8.5cm,clip=t}}
\caption[]{ Same as Fig. 3 for the I, V$-$I color magnitude diagram of
the And~VI dwarf galaxy. 
The right panel also shows the mean RGB lines of five MWG globular
clusters from da Costa et al. (1990) ranging in metallicity from very
low values (M15) to high ones (47 Tuc) as indicated. The RGB's of the
MWG globulars have been shifted to a true distance modulus of 24.5.}
\end{figure}

Figure 3 shows the V, B$-$V color magnitude diagram (CMD) and Figure 4 the 
V, V$-$I CMD. They are already corrected for a foreground reddening of
0.15 mag in B which was derived from Burstein \& Heiles (1978). Only
stars with a photometric accuracy better than 0.15 in both bands are
shown. 

The right panels of Fig. 3 and 4 show only those stars with distances
from the center of the And~VI dwarf galaxy smaller than 175 arc
sec. The left panels represent the stars from the rest of the CCD
frames, a field about 3.5 times larger. A careful comparison indicates
that essentially all stars brighter than I$\sim$20.5 (Fig. 4) and
V$\sim$21.5 (Fig. 3) are MWG stars. Also, any faint and blue
stars with V$-$I$<$0.6 (Fig. 4) either belong to the MWG, or they 
are unresolved background galaxies.

Stars which are fainter than I = 20.5 and redder than V$-$I = 0.6 (V =
22.0, B$-$V = 0.3) are much more abundant within the area encompassed by
the And~VI dwarf galaxy than they are in the field. In particular,
a strongly populated clump of stars is visible around V$-$I$\sim$1.0 in
the right panel of Fig. 4. An analogous feature is present in the
right panel of Fig. 3.  These are the brightest stars of the And~VI
dwarf galaxy.

These brightest stars could be - according to their colors - either
red supergiants and intermediate aged AGB stars or old RGB and AGB
stars. As there is no evidence for recent star formation, neither in our
images (no knotty distribution of the stars, unresolved clumps) nor in
the CMDs (no blue and yellow supergiants which are usually brighter
and easier to detect, no blue blume), these stars are probably RGB
stars (and some AGB stars) of an old population. Indeed, a comparison
with the RBG tracks of galactic globular clusters strongly supports
this identification (Figs. 3 and 4). This is further strengthen by the fact
that Armandroff et al. (1999) did not detect H$\alpha$. This
comparison also indicates that the RGB of the And~VI dwarf galaxy
seems to be a mixed bag of metallicities, with most of the stars
occupying the region of the low metallicity tracks. Our CMDs also
resemble those found for other dSph companions of M31 (see Armandroff
et al., 1998, and references therein).

A clear discontinuity in population density is visible in the V,V$-$I
CMD at I = 20.5. We did I-band star counts inside and outside the area of
the And~VI dwarf galaxy in a color range of 0.6 $<$V$-$I$<$ 2.5. The field
number counts were used to statistically correct the counts inside the
dwarf galaxy for the foreground contamination. The resulting
luminosity function is shown in Fig. 5. The I-band number counts show a
strong step at I = 20.5$\pm$0.2. We identify this feature with the
TRGB. As we are dealing with a low-metallicity system (see above), we
can apply the calibration of Lee et al. (1993), thus, M(I)$_{\rm TRGB} =
-4.0$. This yields a true distance modules of 24.5$\pm$0.20$\pm$0.15.
The first error gives the measurement error as derived from the
location of the TRBG above, while the second error is the systematic
error which was estimated following Schulte-Ladbeck et al. (1998).
The leading error in our case is the accuraccy with which we can
derive the TRGB magnitude in the I counts.

\begin{figure} 
\centerline{\psfig{figure=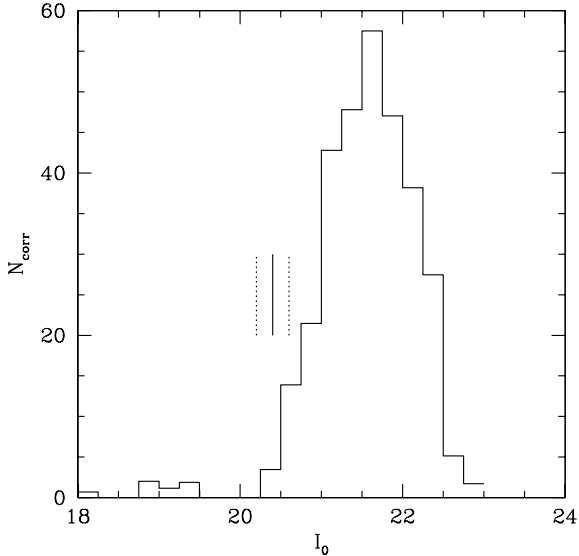,width=8.5cm,clip=t}}
\caption[]{Foreground-corrected I-band number counts of And~VI 
RGB and AGB stars (color range 0.6 $<$ V$-$I $<$ 2.5). The horizontal bar
indicates the value and errors of the TRGB.} 
\end{figure}

The I-band number counts shown in Fig. 5 are still to be corrected for
incompleteness which normally set in about 2 magnitudes above the
observational limits. Thus, while the counts shown should not be used
to construct luminosity functions, the TRGB brightness is still well
above the regime of severe incompleteness.

\section{Conclusions}

We resolved a red population of stars in the newly found And~VI dwarf
galaxy, also designated Pegasus dwarf. This \lq red tangle\rq\ is most
probably composed of red giant stars, plus some stars from an old
asymptotic giant branch. The population seems to be old and
metal-poor. Applying the TRGB method, we derived a true distance
modulus which places And~VI at the distance of M31. Thus, the total
magnitude turns into a rather low absolute one, M$_{V,0} = -10.4\pm0.2$
only!  This is one of the faintest galaxies found so far. The metric
value of the scale length of the surface brightness distribution is
$305\pm11$ pc. The structural parameters and the resolved population
indicate that And~VI is a dSph galaxy.

The newly derived distance modulus suggests that 
And~VI is a companion of M31. The projected distance between M31 and
the And~VI dwarf galaxy is about 271 kpc, as large as that of the most
distant dSph companion of the MWG (Leo I, Mateo, 1998). And~VI is
situated in a region which so far was free of known M31 satellites
(see Fig. 2 of Armandroff et al, 1998). The question has arisen
whether or not the M31 dSph system is different from the MWG, which is
populated by at least 9 dSphs.  The detection of two new dSph galaxies
associated with M31, namely And~V (Armandroff et al., 1998) and And~VI
(Karachentsev \& Karachentseva, 1998, Armandroff et al. 1999, this
work) now brings the census of M31 dSph companions up to 5, somewhat
more in line with what has been observed in the vicinity of the
MWG. This suggests that any apparent difference between the M31 and
the MWG dSph systems is probably due to incompleteness.  We derive a
very faint absolute V magnitude for And~VI, confirming that dSphs as
faint as Draco and Ursa Minor exist around M31, as previously
suspected (Armandroff et al. 98).

\acknowledgements
We like to thank the Calar Alto staff for his kind assistance during the
observations. We were supported by the DFG (hopp/1801-1) and by the
SFB 375, and by GO-07859.01-96A (to rsl).

{}

\begin{thebibliography}{}


\bibitem{}
Armandroff, T.E., Davies, J.E., Jacoby, G., 1998, AJ Nov. 1998 in print

\bibitem{}
Armandroff, T.E., Davies, J.E., Jacoby, G., 1999, in preparation, see 
http://aloe.tuc.noao.edu/jacoby/dwarfs.html (IAU Coll. 171, Cardiff)

\bibitem{}
Bender, R., M\"ollenhoff, C., 1987, A\&A 177, 71


\bibitem{} 
Burstein, D., Heiles, C., 1978, AJ 225, 40

\bibitem{}
Da Costa, G.S., Armandroff, T.E., 1990, AJ 100, 162

\bibitem{}
Durrel, P.R., Harris, W.E., 1993, AJ 105, 1420

\bibitem{}
Hesser, J.E., Harris, W.E., Vandenbergh, D.A., Allwright, J.W.B.,
Shott, P.S., Stetson, P.B., 1987 PASP 99, 739

\bibitem{} 
Karachentsev, I.D., Karachentseva, V., 1998, Dwarf Tales, 3, 1

\bibitem{}
Landolt, A.U., 1992, AJ 104, 340

\bibitem{}
Lee, M.G., Freedman, W.L., Madore, B.F., 1993 ApJ 417, 553 

\bibitem{}
Mateo, M., 1998, ARA\&A 36, 435

\bibitem{}
Schulte-Ladbeck, R.E., Crone, M.M., Hopp, U., 1998, ApJ 493, L23

\bibitem{}
Stetson, P., 1987, PASP 99, 191

\bibitem{}
van den Bergh, S., 1972, ApJ 171, L31

\end{thebibliography}
\end{document}